\documentclass[letterpaper,table,xcdraw]{mc2025}

\usepackage{todonotes}
\usepackage[caption=false]{subfig}
\usepackage{multirow}
\usepackage{xcolor}

\usepackage{nomencl} 
\makenomenclature

\title{
Derivative Source Method for Monte Carlo Transport Calculation of\\
Sensitivities to Material Densities and Dimensions
}

\addAuthor[variansi@oregonstate.edu]{Ilham Variansyah}{1}
\addAuthor{Ryan G. McClarren}{2}
\addAuthor{Todd S. Palmer}{1}

\addAffiliation{1}{Nuclear Science and Engineering, Oregon State University, Corvallis, Oregon}
\addAffiliation{2}{Aerospace and Mechanical Engineering, University of Notre Dame, Notre Dame, Indiana}

\Abstract{%
The Derivative Source Method (DSM) takes derivatives of a particle transport equation with respect to selected parameters and solves them via the standard Monte Carlo random walk simulation along with the original transport problem.
The Monte Carlo solutions of the derivative equations make the sensitivities of quantities of interest to the selected parameters.
In this paper, we show that DSM can be embedded in Monte Carlo simulation to simultaneously calculate sensitivities of transport phase-space solution to multiple object dimensions and material densities.
We verify and assess the efficiency of DSM by solving a multigroup neutronic system of a source-driven fuel-moderator-absorber slab lattice and calculating the fast and slow flux sensitivity coefficient distributions to the fuel and absorber dimensions, as well as the fuel, moderator, and absorber densities.
The results are compared to those obtained from conventional finite difference sensitivity calculations.
A figure of merit, defined as the product inverse of runtime and square of relative error, is used to assess method efficiencies.
For a couple of the calculated sensitivities, well-configured finite difference calculations are the most efficient, followed closely by DSM.
However, since seeking such well-configured finite difference is not always practical, in the rest of the cases, including the all-parameter simultaneous sensitivity calculation, DSM has the highest efficiency, demonstrating its robustness.
}


\shortTitle{MC Derivative Source Method for sensitivity calculation}

\authorHead{Ilham Variansyah, et al.}

\begin{document}

\section{Introduction}\label{sec:intro}

Calculating the local sensitivity of a quantity of interest is a valuable task in various fields~\cite{McClarren2018UQBook}.
In computational particle transport, the motivation behind calculating the sensitivity lies in gaining a quantitative understanding of how small changes around nominal values of input parameters, such as model geometry and material composition of a system, affect the resulting particle phase-space distribution and its derived quantities, such as global parameters and particle streaming and interaction rates in regions of interest.
Applications of local sensitivity analysis generally include design optimizations, identifying influential input parameters (at the nominal values) for further analyses, evaluations or data assimilation, perturbation analysis, and estimating output variance given distributions of uncertain parameters.

Recently proposed by Yamamoto and Sakamoto~\cite{YamamotoJCP2022}, the Monte Carlo sensitivity calculation method Derivative Source Method (DSM) takes the derivative of the transport equation and solves it via the standard Monte Carlo random walk simulation along with the original transport problem.
The Monte Carlo solution of the derivative equation makes the sensitivity of the quantity of interest.
Derivative source particles are sampled during the forward transport simulation and are transported along with the forward physical particles following the same standard random walk but instead score into their respective sensitivity tally bins.
DSM offers exact Monte Carlo estimation (free of approximation or discretization), does not require adjoint calculation, and is relatively straightforward to extend to calculate high-order sensitivities, including the cross-parameter terms (mixed partial derivatives).

Yamamoto and Sakamoto demonstrated DSM application to calculate first-order and second-order sensitivities of particle surface fluxes to material cross-section data (capture, scattering, and scattering anisotropy factor) of a specified region in a system.
In this work, we apply DSM to calculate sensitivities to object dimensions and material densities.
Furthermore, we are interested in calculating the sensitivities of the global solution---i.e., the particle phase-space flux distribution throughout the system---binned into a fine tally mesh to emphasize the benefit of DSM as a non-adjoint method.
With the standard adjoint method~\cite{McClarren2018UQBook}, one would need to run an adjoint problem for each tally mesh bin, making it prohibitively expensive.
Nevertheless, adjoint methods are highly efficient in calculating sensitivities of key integral quantities in a few localized regions (e.g., detectors) and the entire domain (e.g., neutronics multiplication factor) to numerous parameters, even nuclide-wise energy-dependent cross-section data~\cite{KiedrowskiNSE2017, PerfettiNSE2016}.
In contrast, DSM is most efficient for calculating sensitivities of numerous, and even phase-space, quantities to a few key parameters.
Therefore, DSM can be seen as a good complement to the adjoint methods in the sensitivity analysis toolkit.

The rest of the paper is organized as follows.
In \Cref{sec:dsm}, we present the derivation of DSM for calculating flux solution sensitivities to object dimensions and material densities. Then, in \Cref{sec:dsm_mc}, we discuss how to embed DSM into a Monte Carlo transport simulation.
While the method is formulated in a simple one-dimensional transport framework, generalization is straightforward and discussed.
In \Cref{sec:result}, we devise a test problem and present a verification of the proposed DSM Monte Carlo implementation.
Results with conventional finite difference methods are also presented and compared to those of DSM to assess their relative efficiencies.
Finally, \Cref{sec:summary} summarizes the key findings and discusses future work.

\section{Derivative Source Method (DSM)} \label{sec:dsm}

Let us consider a simple yet illustrative mono-energetic, one-dimensional slab steady-state problem with scattering:
\begin{equation}
    \mu \frac{\partial \psi}{\partial x}
    +
    \Sigma_t(x) \psi(x,\mu)
    =
    \int_{-1}^1 
        \Sigma_s (x,\mu' \rightarrow \mu) \psi(x,\mu')
    d\mu'    
    +
    q(x,\mu),
    \quad x \in [0,X], 
    \quad \mu \in [-1,1],
    \label{eq:nte}
\end{equation}
with specified boundary conditions $\psi(0,\mu>0)=\psi_L(\mu)$ and $\psi(X,\mu<0)=\psi_R(\mu)$.
We use symbols typically used in the neutron transport equation~\cite{Duderstadt1976NRA}: $\psi(x,\mu)$ is particle angular flux, $q(x,\mu)$ is independent particle source, $\psi_L(\mu>0)$ and $\psi_R(\mu<0)$ are the boundary conditions, and $\Sigma_t(x)$ and $\Sigma_s(x,\mu' \rightarrow \mu)$ are respectively material total and differential scattering reaction cross-sections.
Now, let us assume that the system has two regions with single-nuclide materials; the cross-sections of the resulting problem can be piece-wisely expressed by the Heaviside function $H(x)$:
\begin{equation}
    \Sigma_t(x)
    =
    \Sigma_{t,1} H(l-x)
    +
    \Sigma_{t,2} H(x-l),
    \label{eq:totalXS_heaviside}
\end{equation}
\begin{equation}
    \Sigma_s(x,\mu \rightarrow \mu)
    =
    \Sigma_{s,1}(\mu' \rightarrow \mu) H(l-x)
    +
    \Sigma_{s,2}(\mu' \rightarrow \mu) H(x-l),
    \label{eq:scatteringXS_heaviside}
\end{equation}
\begin{equation}
    \Sigma_{t,i}
    =
    N_i \sigma_{t,i},
    \quad i = 1, 2,
\end{equation}
\begin{equation}
    \Sigma_{s,i} (\mu' \rightarrow \mu)
    =
    N_i \sigma_{s,i}(\mu' \rightarrow \mu),
    \quad i = 1, 2,
\end{equation}
where $l$ is the position of the intersection surface and $N_i$, $\sigma_{t,i}$, $\sigma_{s,i}(\mu' \rightarrow \mu)$ are respectively atomic density and total and differential scattering microscopic cross-sections of material $i$.

Our goal is to get sensitivities of particle flux $\phi(x)=\int_{-1}^1\psi(x,\mu) d\mu$ with respect to the surface position $l$ and the atomic densities $N_1$ and $N_2$:
\begin{equation}
    \hat{\psi}(x,\mu)
    =
    \frac{\partial \psi}{\partial l}(x,\mu),
    \quad
    \tilde{\psi}_i(x,\mu)
    =
    \frac{\partial \psi}{\partial N_i}(x,\mu),
    \quad i=1, 2.
\end{equation}
We use accents hat $\hat{(\cdot)}$ and tilde $\tilde{(\cdot)}$ to denote quantities associated with the surface position and material atomic density sensitivity parameters, respectively.
We note that the surface position $l$ determines the thicknesses, or dimensions, of the two slab objects in the system.
Thus, this type of sensitivity parameter (surface position) hereafter will be referred to as ``object dimensions".
Furthermore, since we consider single-nuclide materials in this illustrative model, we can get away from the technicalities of different types of densities of material.
Therefore, this type of sensitivity parameter (material atomic density) hereafter will be referred to as ``material densities".
Next, we formulate DSM for solving $\hat{\psi}(x,\mu)$ and $\tilde{\psi}_i(x,\mu)$.

\subsection{Sensitivity to Object Dimension}

Taking the derivative of \Cref{eq:nte} with respect to the surface position $l$:
\begin{equation}
    \mu \frac{\partial \hat{\psi}}{\partial x}
    + \Sigma_t(x) \hat{\psi}(x,\mu)
    =
    \int_{-1}^1 
        \Sigma_s (x,\mu' \rightarrow \mu) \hat{\psi}(x,\mu')
    d\mu'    
    + \hat{q}(x,\mu),
    \label{eq:nte_surface}
\end{equation}
where
\begin{equation}
    \hat{q}(x,\mu)
    =
    -\hat{\Sigma}_t(x) \psi(x,\mu)
    + 
    \int_{-1}^1 
        \hat{\Sigma}_s (x,\mu' \rightarrow \mu) \psi(x,\mu')
    d\mu'.
    \label{eq:q_surface_}
\end{equation}
Here, we assume that $q(x,\mu)$ is independent of $l$.
\Cref{eq:nte_surface} is almost identical to \Cref{eq:nte}, except that \Cref{eq:nte_surface} solves for the derivative particle $\hat{\psi}(x,\mu)$ instead of the phyiscal particle $\psi(x,\mu)$.
In other words, if we know the derivative source $\hat{q}(x,\mu)$, we can solve \Cref{eq:nte_surface} for $\hat{\psi}(x,\mu)$ in the same way as we solve the standard transport problem \Cref{eq:nte} for $\psi(x,\mu)$.

Per \Cref{eq:totalXS_heaviside} and \Cref{eq:scatteringXS_heaviside}, and since $dH(x)/dx=\delta(x)$, we can evaluate the derivatives of the cross-sections~\cite{KiedrowskiICNC2011}:
\begin{equation}
\begin{split}    
    \hat{\Sigma}_t(x)
    &=
    \frac{\partial \Sigma_t}{\partial l}(x)
    \\
    &=
    (\Sigma_{t,1} - \Sigma_{t,2}) \delta(x-l),
    \label{eq:totalXS_hat}
\end{split}
\end{equation}
\begin{equation}
\begin{split}
    \hat{\Sigma}_s(x,\mu' \rightarrow \mu)
    &=
    \frac{\partial \Sigma_s}{\partial l}(x,\mu' \rightarrow \mu)
    \\
    &=
    \left[
        \Sigma_{s,1} (\mu' \rightarrow \mu)
        -
        \Sigma_{s,2} (\mu' \rightarrow \mu)
    \right] \delta(x-l),
    \label{eq:scatteringXS_hat}
\end{split}
\end{equation}
where $\delta(x)$ is the Dirac delta function and $\delta(x) = \delta(-x)$.
Therefore, substituting \Cref{eq:totalXS_hat} and \Cref{eq:scatteringXS_hat} into \Cref{eq:q_surface_}, the derivative source can be rewritten as
\begin{align}
\begin{split}
    \hat{q}(x,\mu)
    =
    \left\{
        -\left(
            \Sigma_{t,1} - \Sigma_{t,2}
        \right)
        \psi(x,\mu)
        +
        \int_{-1}^1 
            \Sigma_{s,1} (\mu' \rightarrow \mu) \psi(x,\mu')
        d\mu'    
    \right.
    \\
    \left.
        -
        \int_{-1}^1 
            \Sigma_{s,2} (\mu' \rightarrow \mu) \psi(x,\mu')
        d\mu'    
    \right\}
    \delta(x-l).
    \label{eq:q_surface}
\end{split}
\end{align}
\Cref{eq:q_surface} indicates that we need $\psi(l,\mu)$ to determine $\hat{q}(x,\mu)$, so that we can solve for $\hat{\psi}(x,\mu)$ per \Cref{eq:nte_surface}, whose operator is equivalent to \Cref{eq:nte} for solving $\psi(x,\mu)$.

\Cref{eq:nte_surface} and \Cref{eq:q_surface} can be generalized into problems with multiple, say $J$, sensitivity surfaces, such that we are solving for $\hat{\psi}_j(x,\mu)$ given $\hat{q}_j(x,\mu)$ and the cross-section notations can be replaced with $\Sigma_{t,j}^-$ and $\Sigma_{t,j}^+$, denoting the cross-sections of the materials on the outward-negative and -positive sides of the surface $j$ defined by the position $l_j$.
This generalized equation is shown in \Cref{eq:q_surface_general}.

\subsection{Sensitivity to Material Density}

Taking the derivative of \Cref{eq:nte} with respect to the material atomic density $N_i$, we get
\begin{equation}
    \mu \frac{\partial \tilde{\psi}_i}{\partial x}
    + \Sigma_t(x) \tilde{\psi}_i(x,\mu)
    =
    \int_{-1}^1 
        \Sigma_s (\mu' \rightarrow \mu) \tilde{\psi}_i(x,\mu')
    d\mu'    
    + \tilde{q}_i(x,\mu),
    \label{eq:nte_material}
\end{equation}
which is similar to \Cref{eq:nte_surface}, except that the derivative source is defined as
\begin{equation}
    \tilde{q}_i(x,\mu)
    =
    \left\{
        -\sigma_{t,i} \psi(x,\mu)
        +
        \int_{-1}^1 
            \sigma_{s,i} (\mu' \rightarrow \mu) \psi(x,\mu')
        d\mu'    
    \right\}
    \delta(x \in V_i),
    \label{eq:q_material}
\end{equation}
where, per \Cref{eq:totalXS_heaviside} and \Cref{eq:scatteringXS_heaviside}, $V_1=[0,l]$ and $V_2=[l,X]$, and $\delta(x \in V_i)$ returns one if the position $x$ is in the region of the material $i$ and zero otherwise.
Here, we assume that $q(x,\mu)$ is independent of $N_i$.
\Cref{eq:q_material} indicates that we need $\psi(x\in V_i,\mu)$ to determine $\tilde{q}_i(x,\mu)$, so that we can solve for $\tilde{\psi}_i(x,\mu)$ per \Cref{eq:nte_material}, whose operator is equivalent to \Cref{eq:nte} for solving $\psi(x,\mu)$.
Note that instead of just two, we may have an arbitrary number of sensitivity materials $I$.

\section{Embedding DSM into Monte Carlo Simulation} \label{sec:dsm_mc}

We note that DSM formulation in \Cref{sec:dsm} is applicable for both deterministic and Monte Carlo transport methods.
In this section, we present a way to embed DSM into the standard Monte Carlo forward transport simulation.
We also discuss how the DSM formulation can be extended to multi-dimensional (with general quadric interfaces) continuous-energy transport problems.

Based on \Cref{eq:nte}, \Cref{eq:nte_surface}, and \Cref{eq:nte_material}, we can simultaneously transport the physical particle $\psi(x,\mu)$ along with the ``imaginary" derivative particles $\hat{\psi}_j(x,\mu)$ and $\tilde{\psi}_i(x,\mu)$ using the same particle random walk Monte Carlo algorithm.
However, tally scores made by the derivative particles accumulate to their respective sensitivity estimates.
This can be implemented by introducing an attribute, say ``$\text{sensitivity\_ID}$", to the particle data structure and adding an additional outer dimension of size $1+N_p$ to the tally array, where $N_p=I+J$ is the number of sensitivity parameters.
The particle attribute $\text{sensitivity\_ID}$ corresponds to the outermost tally index, where $\text{sensitivity\_ID}=0$ corresponds to the tally estimates of the quantities of interest and $\text{sensitivity\_ID}>0$ corresponds to their sensitivities.
Note that we essentially group $\psi(x,\mu)$, $\hat{\psi}_j(x,\mu)$, and $\tilde{\psi}_i(x,\mu)$ into sensitivity groups, which are treated similarly as neutron energy groups in a multigroup simulation.
Another consideration in the implementation is to anticipate negative weights from the derivative particles; for example, in a technique that requires comparing particle weight against a threshold value (such as in weight roulette), we should neglect the particle weight sign by taking its absolute value instead.

The last modification needed to embed DSM into the standard Monte Carlo simulation is to introduce sampling procedures for the derivative sources $\hat{q}_j(x,\mu)$ and $\tilde{q}_i(x,\mu)$.
The following two subsections discuss how the derivative source particles can be sampled based on surface-crossing and collision events of the physical particle.
Then, we discuss general remarks on the integration of the proposed derivative source sampling procedure into a Monte Carlo code.

\subsection{Sampling Derivative Sources for Sensitivity to Object Dimension}

The generalized forms of \Cref{eq:nte_surface} and \Cref{eq:q_surface} are
\begin{equation}
    \mu \frac{\partial \hat{\psi}_j}{\partial x}
    + \Sigma_t(x) \hat{\psi}_j(x,\mu)
    =
    \int_{-1}^1 
        \Sigma_s (x,\mu' \rightarrow \mu) \hat{\psi}_j(x,\mu')
    d\mu'    
    + \hat{q}_j(x,\mu),
    \label{eq:nte_surface_general}
\end{equation}
\begin{align}
\begin{split}
    \hat{q}_j(x,\mu)
    =
    \left\{
        -\left(
            \Sigma_{t,j}^- - \Sigma_{t,j}^+
        \right)
        \psi(x,\mu)
        +
        \int_{-1}^1 
            \Sigma_{s,j}^- (\mu' \rightarrow \mu) \psi(x,\mu')
        d\mu'    
    \right.
    \\
    \left.
        -
        \int_{-1}^1 
            \Sigma_{s,j}^+ (\mu' \rightarrow \mu) \psi(x,\mu')
        d\mu'    
    \right\}
    \delta(x-l_j).
    \label{eq:q_surface_general}
\end{split}
\end{align}
The Dirac delta function in \Cref{eq:q_surface_general} indicates that to sample from the derivative source $\hat{q}_j(x,\mu)$, we need an estimate of $\psi(x, l_j)$, which can be obtained via the surface-crossing estimator~\cite{Dupree2022MC}.
This means that whenever the physical particle crosses the surface $j$, we can get a sample of the derivative source particle.
However, there are three terms in \Cref{eq:q_surface_general}: (1) collision, (2) negative-side scattering, and (3) positive-side scattering.
One method for sampling the derivative source is to produce one source particle for each of the terms in the equation (which is implemented to some extent in ~\cite{KiedrowskiICNC2011}).
With this method, three derivative source particles are produced whenever a physical particle crosses the surface.

The source particle from the collision term in \Cref{eq:q_surface_general} copies the phase space of the physical particle; this includes the direction vector and energy in 3D continuous-energy problems.
The source particle is then assigned the weight of $-(\Sigma_{t,j}^- - \Sigma_{t,j}^+) (w/|\mu|)$, where $w$ and $\mu$ are the weight and cosine direction of the physical particle crossing the surface.
Note that the $w/|\mu|$ is the surface-crossing estimator for particle flux~\cite{Dupree2022MC}.
In the case of a particle crossing a more general quadric surface in 3D problems, the particle cosine angle $\mu$ is replaced by the cosine angle between the surface's outward normal and the particle directions.

As for the source particles from the scattering terms in \Cref{eq:q_surface_general}, their phase spaces are sampled as if they come out from scattering events of the physical particle with the material in the negative and positive sides of the surface, respectively.
In general, these would involve sampling the scattering nuclide and its continuous-energy scattering law.
Finally, the source particles are then assigned the weight of $\Sigma_{s,j}^- (w/|\mu|)$ and $-\Sigma_{s,j}^+ (w/|\mu|)$, respectively.
Note that $\Sigma_{s,j}^\pm$ is the total scattering cross-section---i.e., the total integral of the differential one $\Sigma_{s,j}^\pm (\mu' \rightarrow \mu)$.

\subsection{Sampling Derivative Sources for Sensitivity to Material Density}

\Cref{eq:q_material} indicates that we can sample the derivative source particles $\tilde{q}_i(x,\mu)$ whenever the physical particle collides with the material $i$.
There are two terms in $\tilde{q}_i(x,\mu)$: collision and scattering.

The source particle from the collision term copies the phase space of the physical particle and is assigned the weight of $-\sigma_{t,i} (w/\Sigma_{t,i})$, where $(w/\Sigma_{t,i})$ comes from the collision estimator of the flux.
Meanwhile, the source particle from the scattering term is assigned the weight of $\sigma_{s,i} (w/\Sigma_{t,i})$ and its phase space is sampled as if it comes out from a scattering event of the physical particle with the material.
Note that $\sigma_{s,i}$ is the total microscopic scattering cross-section---i.e., the total integral of the differential one $\sigma_{s,i} (\mu' \rightarrow \mu)$.

We note that the same methodology is applicable if the material has multiple nuclides.
In this case, we will sample two source particles for each nuclide comprising the material, and the nuclides' respective atomic ratios would need to be factored into the assigned weight.
A similar treatment is applicable if the selected sensitivity parameter is only a certain nuclide in the material.
In this case, we only need to generate source particles for that particular nuclide.

\subsection{Remarks on the Derivative Source Sampling}
\label{sec:dsm-mc-remark}

\textbf{Managing Particle Bank.}
As discussed in the previous subsections, derivative source particles are produced anytime a physical particle crosses a sensitivity surface or collides with a sensitivity material.
To avoid storing a growing derivative particle population, it is recommended to pause the random walk of the physical particle whenever a derivative particle is produced and run the random walk of the derivative particle and its secondaries until completion.
This can be achieved by temporarily storing the physical particle in a particle bank.
If the particle bank works on a last-in-first-out basis, the physical particle and the derivative particles (including the secondaries) can be automatically managed in the same particle bank.

\textbf{Generalization to Multiplying Media.}
The derivative source sampling techniques described in the previous subsections can be extended to include particle production reactions such as neutron fission~\cite{KiedrowskiICNC2011}.
In this case, five (instead of three) or three (instead of two) derivative source particles are produced whenever a physical particle crosses a sensitivity surface or material, respectively.
Note that the reaction multiplicity is factored in to augment the source particle weights.

\textbf{Alternative Sampling Method for the Derivative Source.}
Given that we produce multiple source particles at each triggering event (surface crossing or collision), one can see a potential issue of excessive derivative source production, leading to a significant increase in runtime.
One alternative to such a method is to produce only one derivative source particle at each of the triggering events.
This is aligned with how reaction type is sampled in analog Monte Carlo---instead of simulating all possible reaction types upon a collision and assigning the secondaries with their respective expected weights, we simulate only one sampled reaction type, and the resulting secondary takes all the weight at play.
This means, for \Cref{eq:q_surface_general}, we produce only one (instead of three) derivative source particle with sampling probability ratios of $|\Sigma_{t,j}^- - \Sigma_{t,j}^+|$, $\Sigma_{s,j}^-$, and $\Sigma_{s,j}^+$ for the source particle to be sampled from the collision, negative-side scattering, or the positive-side scattering term, respectively.
Depending on which source term is sampled, the source particle is assigned the weight of $-\hat{w}_j$, $\hat{w}_j$, or $-\hat{w}_j$, respectively, where
\begin{equation}
    \hat{w}_j
    =
    \left[
        |\Sigma_{t,j}^- - \Sigma_{t,j}^+| 
        +
        \Sigma_{s,j}^-
        +
        \Sigma_{s,j}^+
    \right] \frac{w}{|\mu|}.
\end{equation}

As for \Cref{eq:q_material}, we produce only one (instead of two) derivative source particle with sampling probability ratios of $\sigma_{t,i}$ and $\sigma_{s,i}$ for the source particle to be sampled from the collision or the scattering term, respectively.
Depending on which source term is sampled, the source particle is assigned the weight of $-\tilde{w}_i$ or $\tilde{w}_i$, respectively, where
\begin{equation}
    \tilde{w}_i
    =
    \left[
        \sigma_{t,i}
        +
        \sigma_{s,i}
    \right] \frac{w}{\Sigma_{t,i}}.
\end{equation}
All results presented in this paper follow this alternative sampling approach.

\section{Results and Discussions} \label{sec:result}

\begin{figure}[h!]
    \centering
    \includegraphics[width=0.5\columnwidth]{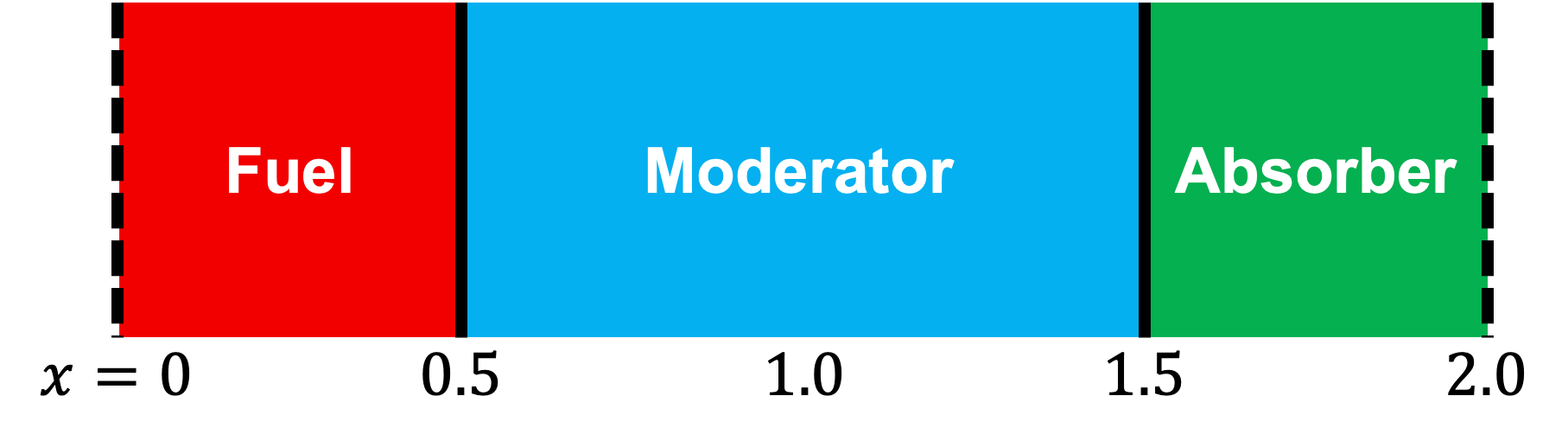}
    \caption{The infinite lattice system of 1-cm-thick slabs of fuel, moderator, and absorber, modeled with reflecting boundaries at $x=0$ and $x=2.0$.}
    \label{fig:lattice_model}
\end{figure}

The proposed DSM Monte Carlo simulation is implemented to the open-source Python-based Monte Carlo code MC/DC~\cite{variansyah_mcdc}.
To verify and assess the efficiency of the method, we devise a test problem of a multigroup infinite slab lattice system that consists of alternating, 1-cm-thick fuel and absorber slabs separated by a 1-cm-thick water moderator.
\Cref{fig:lattice_model} illustrates how the infinite lattice is modeled.
We use the C5G7 material data~\cite{hou2017} for the fuel (UO$_2$), moderator, and absorber (control rod).
Finally, an isotropic uniform neutron source emitting particle at the highest-energy group (1.36 to 10 MeV) is introduced at the center of the water moderator to drive the steady-state subcritical fission chain reaction of the system.
Our quantities of interest are fast (9.2 keV to 10 MeV) and slow (below 9.2 keV) neutron flux distributions, tallied into 200 uniform mesh bins, and the associated sensitivity coefficients with respect to five parameters: (1) fuel thickness represented as the material interface at $x=0.5$, (2) absorber thinness represented as the material interface at $x=1.5$, and (3) fuel, (4) moderator, and (5) absorber densities.
The sensitivity coefficients take the ratios of the sensitivities to the flux solutions, physically representing flux relative changes due to the relative changes in the associated parameters.
All of the MC simulations were run with distributed memory parallelization on the Lawrence Livermore National Laboratory compute platform Quartz, having Intel Xeon E5-2695 v4 CPU architecture with 36 cores per node. Eight compute nodes were used, giving a total of 228 CPU cores.

\begin{figure}[h!]
    \centering
    \includegraphics[width=0.8\columnwidth]{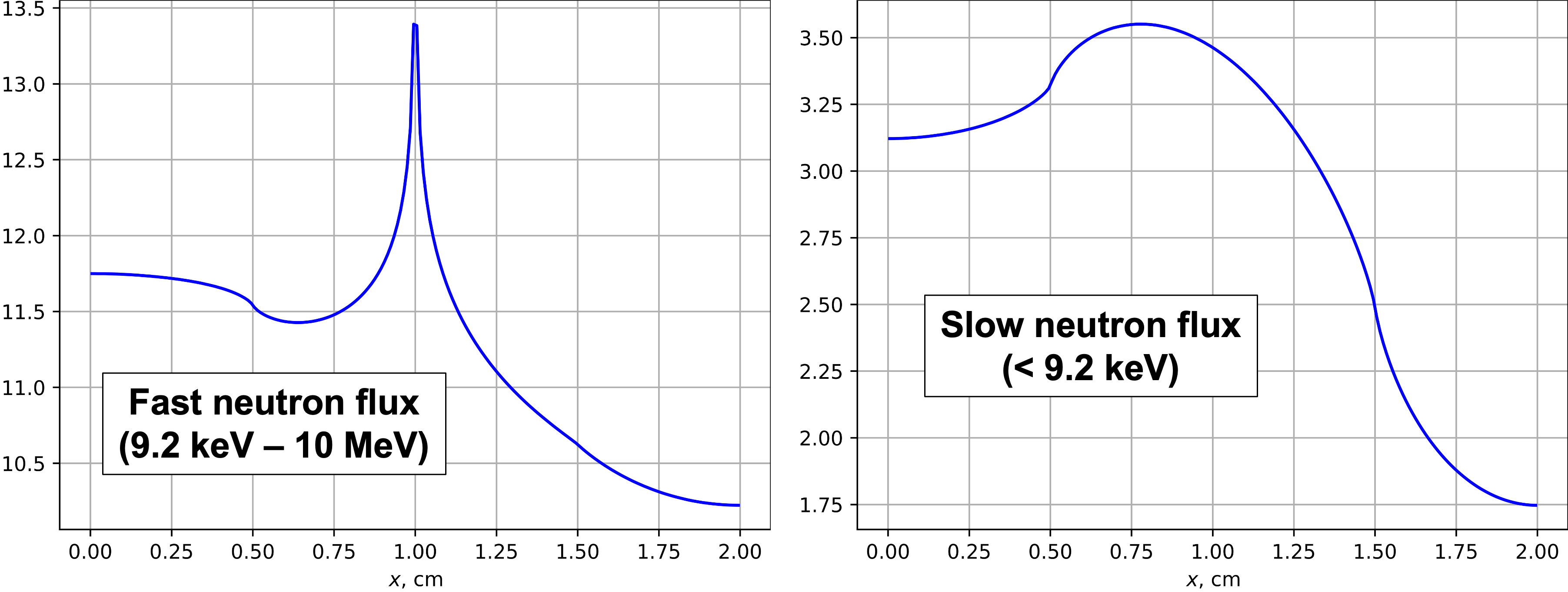}
    \caption{Fast (left) and slow (right) neutron flux solutions for the infinite lattice test problem, run with $10^9$ source particle histories.}
    \label{fig:lattice_solution}
\end{figure}

\begin{figure}[h!]
    \centering
    \includegraphics[width=0.8\columnwidth]{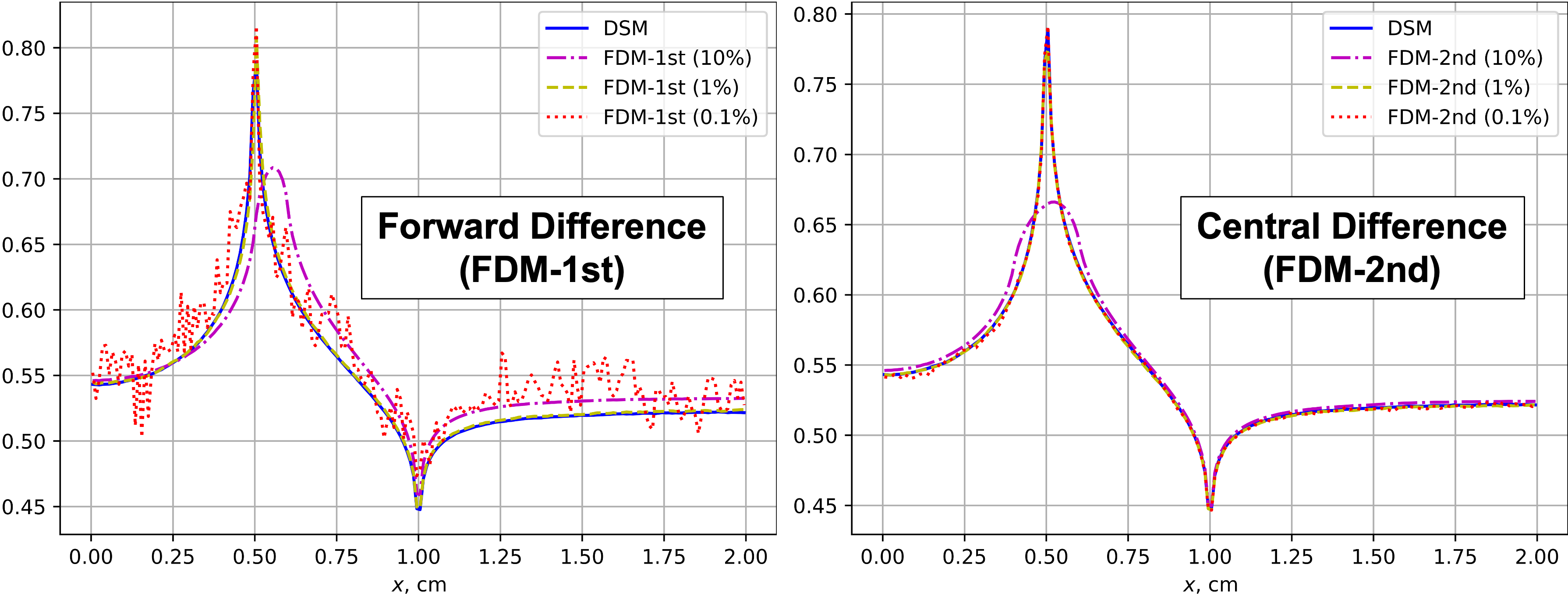}
    \caption{DSM and forward (left) and central (right) FDMs results for the fast flux sensitivity coefficient distribution to the fuel thickness, run with $10^9$ source particle histories. Parentheses on the legends indicate the perturbation of the fuel interface position at $x=0.5$ for the FDMs.}
    \label{fig:lattice_fdm_compare}
\end{figure}

\begin{figure}[hp!]
    \centering
    \includegraphics[width=1.0\columnwidth]{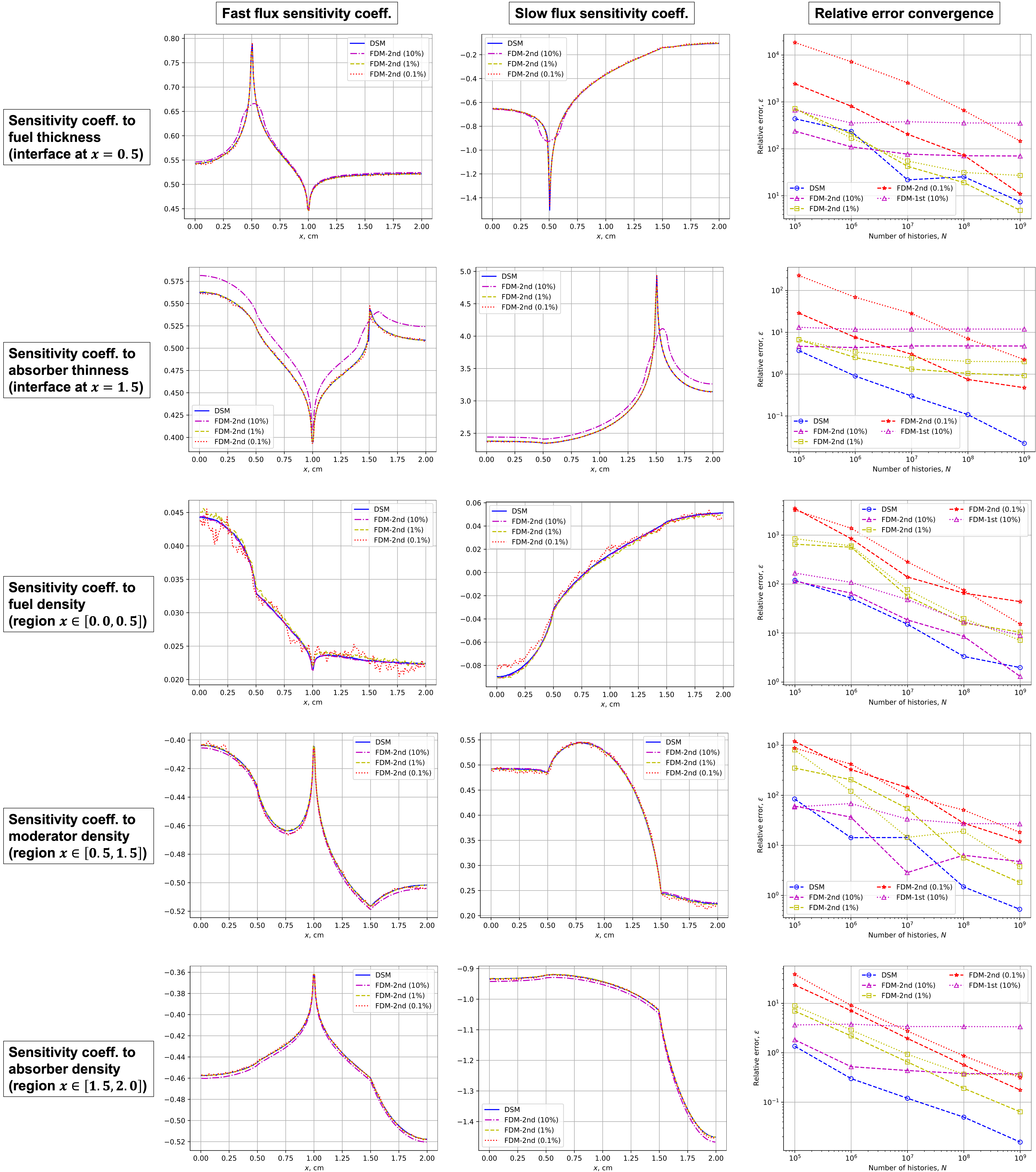}
    \caption{DSM and FDM results for fast (left) and slow (middle) flux sensitivity coefficient distributions, run with $N=10^9$ source particle histories, as well as their overall relative error convergences (right), to the five sensitivity parameters: fuel thickness, absorber thinness, and fuel, moderator, and absorber densities (top to bottom).
    First-order FDM results are not shown in the left and middle figures.}
    \label{fig:lattice_sensitivities}
\end{figure}

We run the test problem with DSM and conventional first-order forward and second-order central finite difference methods (FDMs), with three perturbation levels---10\%, 1\%, and 0.1\%---applied to the sensitivity parameters.
\Cref{fig:lattice_solution} shows the resulting fast and slow neutron flux distributions.
Five sensitivity coefficient distributions are calculated for each of the fast and slow fluxes.
\Cref{fig:lattice_fdm_compare} shows one of them, but particularly, it illustrates how the accuracy of the FDM methods depends on the perturbation level and the method's accuracy order.
Smaller perturbations lead to more accurate but less precise (more stochastic noise) results.
These are clearly evident for the low-order forward FDM (left figure).
On the other hand, the higher-order central FDM (right figure) produces results that are significantly more accurate and precise, converging to the solution of DSM as the perturbation gets smaller.

\Cref{fig:lattice_sensitivities} shows all the 10 sensitivity coefficient distributions (five sensitivity parameters for each of the two-group fluxes).
The results of DSM and FDM are in good agreement.
The FDM results get more accurate and closer to those of the DSM as smaller perturbations are used (particularly see the peaks and dips in the sensitivities to the object dimensions). However, smaller perturbations also lead to more noisy results in some cases.
\Cref{fig:lattice_sensitivities} also shows the convergence of the relative error 2-norm (from all energy groups and spatial mesh bins) with respect to the number of source particle histories $N$.
DSM results with $N=10^{10}$ are used as the reference solution.
The expected $1/\sqrt{N}$ convergences are evident for methods that are run accurately enough.

\Cref{tab:runtime} shows the runtime taken by DSM relative to the time taken for solving the transport problem (without sensitivity calculation) and the forward and central difference methods (all with $N=10^9$).
It is shown that, in most of the cases, DSM runs slower than the two FDMs, except for calculating sensitivities to fuel and absorber densities, in which DSM runs faster than the central FDM.
The significantly higher runtimes for calculating sensitivities to fuel thickness and absorber thinness indicate that the test problem is optically thin such that interface crossing occurs very frequently, causing excessive derivative particle source production.
The higher runtimes for calculating the sensitivity to moderator density can be simply due to the source particle spawning point located at the center of the moderator.

\begin{table}[h!]
    \centering
    \caption{DSM runtimes for calculating sensitivity coefficient distributions to the parameters (individually and simultaneously). The presented runtimes are relative to those of (1) solving the particle transport problem (without sensitivity calculation) and (2) forward and (3) central difference methods.}
    \begin{tabular}{|l|rrr|}
        \hline
        \rowcolor[HTML]{EFEFEF} 
        \cellcolor[HTML]{EFEFEF} & \multicolumn{3}{c|}{\cellcolor[HTML]{EFEFEF}DSM relative runtime to}                     \\ \cline{2-4} 
        \rowcolor[HTML]{EFEFEF} 
        \multirow{-2}{*}{\cellcolor[HTML]{EFEFEF}\begin{tabular}[c]{@{}l@{}}Sensitivity\\ parameter\end{tabular}} &
          \multicolumn{1}{c|}{\cellcolor[HTML]{EFEFEF}\begin{tabular}[c]{@{}c@{}}Transport solve\\ $T_s$\end{tabular}} &
          \multicolumn{1}{c|}{\cellcolor[HTML]{EFEFEF}\begin{tabular}[c]{@{}c@{}}Forward difference\\ $(1+N_p) \times T_s$\end{tabular}} &
          \multicolumn{1}{c|}{\cellcolor[HTML]{EFEFEF}\begin{tabular}[c]{@{}c@{}}Central difference\\ $(1+2N_p) \times T_s$\end{tabular}} \\ \hline
        Fuel density             & \multicolumn{1}{r|}{2.62}           & \multicolumn{1}{r|}{1.31}          & 0.87          \\ \hline
        Moderator density        & \multicolumn{1}{r|}{5.18}           & \multicolumn{1}{r|}{2.59}          & 1.73          \\ \hline
        Absorber density         & \multicolumn{1}{r|}{2.59}           & \multicolumn{1}{r|}{1.3}           & 0.86          \\ \hline
        Fuel thickness           & \multicolumn{1}{r|}{6.67}           & \multicolumn{1}{r|}{3.33}          & 2.00          \\ \hline
        Absorber thinness        & \multicolumn{1}{r|}{6.00}           & \multicolumn{1}{r|}{3.00}          & 2.00          \\ \hline
        \textbf{All ($N_p = 5$)}    & \multicolumn{1}{r|}{\textbf{19.06}} & \multicolumn{1}{r|}{\textbf{3.18}} & \textbf{1.73} \\ \hline
    \end{tabular}
    \label{tab:runtime}
\end{table}

\begin{figure}[h!]
    \centering
    \includegraphics[width=1.0\columnwidth]{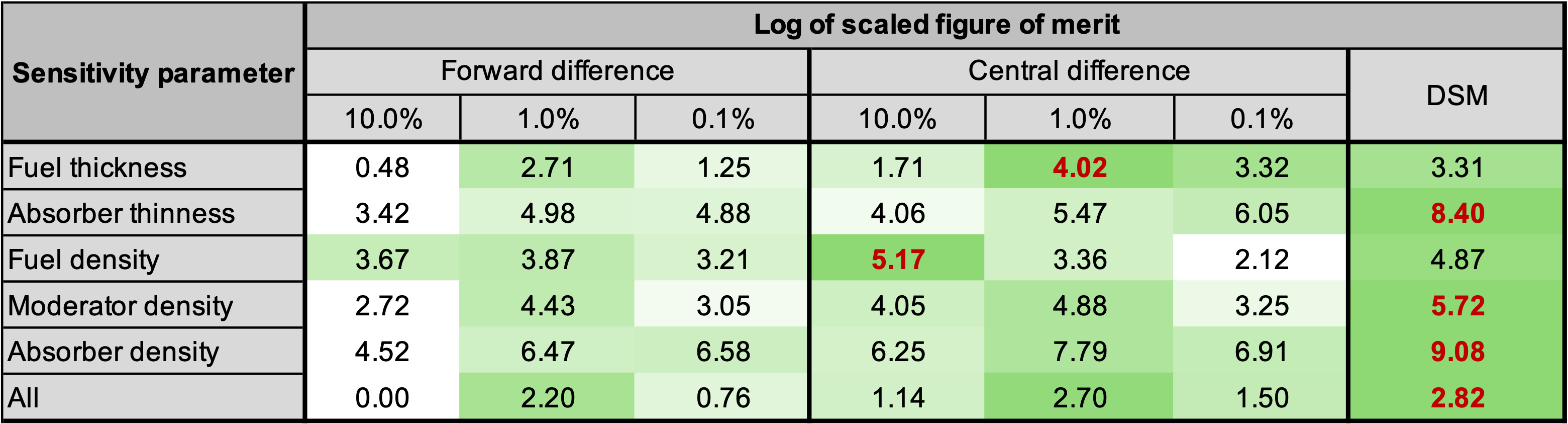}
    \caption{Figure of merit (FOM) map of forward difference (with 10\%, 1\%, and 0.1\% parameter perturbations), central difference, and DSM for calculating the sensitivity coefficient distributions to each individual and all of the sensitivity parameters.
    Bold red-colored values indicate the highest FOMs in their respective rows.
    The green color map indicates the row-wise rank of the FOM.
    All of the simulations are run with $N=10^9$.}
    \label{fig:lattice_fom}
\end{figure}

To properly assess method efficiencies and account for the DSM's superiority in simulation accuracy/precision and inferiority in runtime, we compare the figures of merit (FOM), defined as the product inverse of runtime and square of relative error 2-norm, of all the methods in calculating sensitivities to the individual parameters, as well as all of them altogether.
The resulting FOM map is shown in \Cref{fig:lattice_fom}.
The more conventional definition of FOM that uses the variance of the tally solution is not used as some of the compared methods suffer from inaccuracies to the error from discrete perturbations.
For comparison convenience, we scale the FOMs relative to the lowest value, and then we take its logarithm to the base of 10.
It is found that for calculating sensitivities to fuel thickness and density, central FDM with perturbations of 1\% and 10\%, respectively, are the most efficient, closely followed by DSM.
It seems that we happen to pick well-configured FDMs with the second-order method, as well as the selected perturbations.
However, such a well-configured FDM seems to be not, and possibly can't be (due to the race between increasing accuracy and decreasing precision), obtained in the other cases in which DSM is the most efficient, demonstrating its robustness.

\section{Conclusions and Future Work} \label{sec:summary}

We formulate the derivative source method (DSM) for calculating transport solution sensitivities to material densities and object dimensions.
We then present how the formulated DSM can be embedded in the standard Monte Carlo transport simulation.
We verify and assess the efficiency of DSM by solving a multigroup neutronic system of a source-driven fuel-moderator-absorber slab lattice and calculating the fast and slow flux sensitivity distributions to the fuel and absorber dimensions, as well as the fuel, moderator, and absorber densities.
Based on the figure of merit, well-configured finite difference calculations are the most efficient for a couple of the calculated sensitivities, closely followed by DSM.
However, since seeking such well-configured finite difference is not always practical, in the rest of the cases, including the all-parameter simultaneous calculation, DSM has the highest efficiency, demonstrating its robustness.

Future work includes more formal verification efforts, including comparing against DSM implementations in deterministic methods, as well as testing DSM in more practical, multi-dimensional continuous-energy problems.
\Cref{fig:lattice_fom} shows that while DSM runs faster than the FDM in a few cases, DSM runtime is significantly higher in the rest of the cases.
This is primarily due to the excessive production of derivative source particles, even with the alternative sampling approach that produces only one (not five or three) particles per triggering event (surface crossing or material collision).
DSM can greatly benefit from Monte Carlo techniques that appropriately reduce runtime (and increase precision) by incorporating particle importance, like weight window~\cite{cooper_WW}, to both the physical and ``imaginary'' derivative particles.
It is also interesting to explore DSM applications for calculating second-order sensitivities, including the cross-parameter terms (mixed partial derivatives), particularly those with respect to object dimensions.
Investigating DSM applications for neutronics iterative eigenvalue calculations is important, considering the typical challenges encountered by non-adjoint-based methods in such calculations.
Last but not least, DSM implementation is found to be very similar to the Contributon method~\cite{williams1991generalized} despite significantly different theoretical bases.
Characterizing the connections between the two methods (particularly to the advanced Contributon variant CLUTCH~\cite{PerfettiNSE2016}) would be valuable and may open up improvements and extended applications of both methods.

\section*{Acknowledgements}
This work was supported by the Center for Exascale Monte-Carlo Neutron Transport (CEMeNT), a PSAAP-III project funded by the Department of Energy, grant number DE-NA003967.
The authors thank the reviewers for their constructive comments and suggestions that enhanced the paper and improved future directions of the method development.

\bibliographystyle{mc2025}
\bibliography{main}

\end{document}